\documentclass[12pt]{article}
\usepackage{graphicx}
\usepackage{amsmath}
\usepackage{amsfonts}
\usepackage{color}

\begin{document} \begin{center}
{\huge  Growth kinetics of NaCl crystals in a drying drop of gelatin: transition from faceted to dendritic growth  }\\ \vskip .5cm
Moutushi Dutta Choudhury$^{1a}$, Tapati Dutta$^{2b}$ and Sujata Tarafdar$^{1*}$\\
$^1$ Condensed Matter Physics Research Centre, Physics Department, Jadavpur University, Kolkata 700032, India\\
$^2$ St. Xavier's College, Kolkata 700016, India\\
$^*$Corresponding author, sujata$\_$tarafdar@hotmail.com,
Phone: +91 33 24146666(Ex. 2760), $^a$mou15july@gmail.com, $^b$tapati$\_$mithu@yahoo.com 
\end{center} 
\noindent
Abstract\\ \noindent
We report a study on the kinetics of drying of a droplet of aqueous gelatin containing sodium chloride. The process of drying recorded on video, clearly shows  different regimes of growth leading to a variety of crystalline patterns. Large faceted crystals of $\sim$mm size form in the early stages of evaporation, followed by  highly branched multi-fractal patterns with micron sized features. We simulate the growth using a simple algorithm incorporating aggregation and evaporation, which reproduces the cross-over between the two growth regimes. As evaporation proceeds, voids form in the gel film. The time development of the fluid-void system can be characterized by an Euler number.  A minimum in the Euler number marks the transition between the two regimes of growth.
 \\

 \noindent Keywords: Crystal growth, dendrite, gelatin, sodium chloride, drying droplet

\newpage

\section{Introduction}
The study of droplets of complex fluids, drying through evaporation, reviewed recently by Sefiane \cite{sefiane}  is becoming a rapidly growing field of research. The widespread interest is due to the basic physics aspect as well as promising applications in technology \cite{layani}, quality control and medical diagnostics \cite{solis,brutin,yuri,wheat-grain}. A drying droplet may exhibit a host of features, concentric ring patterns \cite{kaya}, cracks \cite{pauchard,zhang} and other interesting morphology.  
 A fact well known since many years, is that gels provide a good medium for growth of crystals \cite{gel-book}. 

It was shown recently that NaCl added to bio-materials like potato starch, gelatin and carboxy-methyl cellulose crystallizes in different forms \cite{moucolsua, abhracolsua} when a drop of the gel is allowed to dry. The results depend on composition of the droplet, ambient temperature and humidity conditions as well as the nature of the substrate. The kinetics of formation for salt and gel aggregates, observed by  video recording
 \cite{youtube} the drops as they dry, reveals further fascinating features of these systems. In particular, drops of NaCl in gelatin, showed two clearly different regimes of crystalline morphology - (a) large ($\sim$mm) sized regular rectilinear crystals and (b) a fine network-like  multifractal dendritic pattern of typical size $\sim \mu$m. The former grow in the initial stage of drying and the process is slow, while the latter grow subsequently, at a much faster rate.

In the present paper we concentrate on the \textit{growth mechanism} of the crystalline aggregates and in particular on the crossover from the regime (a) to (b). A simulation algorithm is developed, which illustrates the two regimes and the cross-over from one to the other. The model starts with a system of salt-containing gel sites (G), spread over the sample and a seed consisting of crystal (C) sites at the center. The seed crystal grows uniformly along its boundary as long as salt is available.
 As evaporation proceeds, the gel sites may convert to voids (V), where the solvent has evaporated. The distribution of G, C and V sites on the sample is allowed to evolve with time according to rules explained in section (\ref{simulation}). The growth rules are governed by the energy difference due to the event under consideration.

In the previously reported paper \cite{abhracolsua},  a very simple aggregation model was employed to demonstrate a  dendritic multi-fractal \cite{cgd} growth process. The faceted growth was not modelled in this work. The presence of solvent and its evaporation was not considered at all and energy, chemical potential or temperature did not enter the process. So the present approach is a much extended and improved version of the earlier one.

\section{Materials and Method}
\subsection{Sample preparation}
Chemicals used are from Lobachemie, Mumbai. 
 Gelatin in distilled water is the colloidal medium. NaCl (0.03 mol) is
dissolved in 50 ml of water; then 0.5 g of gelatin is mixed with the
solution and stirred for 1 h at 70$^\circ$ C until a homogeneous gel forms.
Drops taken are of  volume 0.5 ml, and the substrate is glass.

\subsection{Observing the desiccation process}
The drop is deposited on a glass cover slip and placed under a microscope (Leica DM750). The drying process is observed and recorded on video.
 Scanning Electron Microscopy (using configuration No. QUO-35357-0614) is also done at Physics Department, Jadavpur University to see the structures at higher resolution. 
 
 To observe the effect of varying humidity, which causes a change in drying rate, a set of samples with same concentration (0.03 M) were dried under different relative humidity conditions RH = 35\% and 45\%.

To observe the effect of varying the salt concentration, samples with salt concentration $c_0$ varying from 0.005 to 0.03 M were dried at the same temperature of 25$^\circ$C.

\subsection{A preliminary `incomplete drying' experiment}
To ascertain the deposition sequence of underlying layers, which are not visible after complete drying, the following experiments are done.

Drops are deposited and allowed to dry for a certain  time interval, before the sample dries fully and the glass surface is washed off lightly with distilled water, leaving the material deposited so far. EDAX is done on this sample, to determine which species have deposited already. For a drying time 1 h, there is hardly any deposition, but for a time interval of 2 h a thin layer is formed, though macroscopic aggregates are not visible. EDAX was done at S.N. Bose National Center for Basic Sciences, Kolkata(JEOL JSM-6360).

\section{Results}
\subsection{The two  regimes of crystal growth}
Both the processes (a) slow growth of large crystals and (b) fast growth of dendrites have been recorded on video. Figures (\ref{facet-gr}) and (\ref{dendrite-gr}) show a few successive stills taken during growth by process (a) and (b) respectively. 

 In the initial stages the system  can be approximately considered as a section of a sphere. 
 Due to the three-dimensional geometry, observing the drying drop  through the microscope has problems. Only one plane can be accurately focussed at a time. Sometimes bubbles and small crystallites float on the upper surface and other structures, though not well focussed, can be made out in regions closer to the substrate. But we can consider a simplified quasi-two-dimensional picture. This picture comes closer and closer to reality with time. The scenario is the following.

Once a rectilinear seed crystal is formed, the growth front proceeds normally to the smooth linear sides, thus the shape of the crystal is more or less preserved during growth, that is the growth is \textit{homothetic} \cite{pimpinelli}. We prefer not to use the term `self-similar' which has been used for this type of growth in \cite{pimpinelli}, as self-similarity is usually attributed to \textit{fractals} , where the same pattern is repeated on all scales. 

\begin{figure}[ht]
\begin{center}
\includegraphics[bb=0cm 0cm 30cm 30cm, scale=0.5, angle=0]{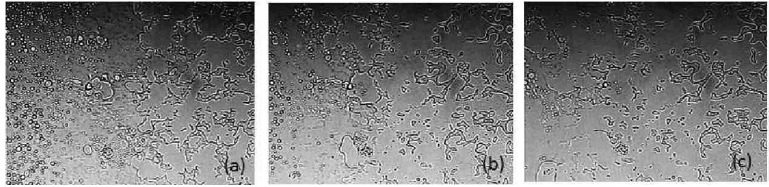}
\end{center}
\caption{(a) - (c) illustrate the void-in-fluid to fluid-in-void transition. The snaps (a), (b), (c) are taken at times 0 s, 1 min. 30 s, 2 min 30 s respectively. The void spaces are invading the area in the photograph from the right. The length of the rectangular panels shown in (a)-(c) is 1.73 mm. } \label{transition}
\end{figure}

This process seems to continue as long as the growing seed crystal is surrounded by the solution on all sides. As evaporation proceeds the layer of salt-gelatin solution thins out and gaps begin to form in it. This is like a phase separation process, with voids and solution aggregates distributed on the surface of the substrate. To start with, the voids are isolated in a sea of salt solution. But as voids increase in size and number, at some point in the evaporation process, the \textit{voids} become connected and a distribution of finite blobs of solution with varying shapes and sizes is seen. The gel is porous and many bubbles can be seen under the microscope. Figures (\ref{transition}a-c) show the bubbly gel on the left side and the growing voids invading from the right as evaporation proceeds.

After this `phase transition' a different growth mechanism (b) starts to dominate over the uniform growth process(a). Near this point the large crystals stop growing uniformly along their periphery as there is no longer a layer of fluid adjacent to the crystal. Subsequent crystallization takes place by a different process (b). From the video \cite{youtube} we see that the isolated blobs of fluid in the surrounding space are attracted to the crystal, particularly the corners and dendritic growth starts from the corners. Figure (\ref{corner}) shows dendrites starting and growing from the corners of a faceted crystal. Initiation of dendritic growth is not completely restricted exclusively to corners of large crystals, they may nucleate at other points in the gel and grow by a similar process. The dendritic growth is very fascinating to watch \cite{youtube}. When a fluid blob reaches the dendrite it forms a square crystallite whose size depends on the size of the liquid blob. A chain of such crystallites attached corner-to-corner proceeds linearly, intermittently branching out at right angles as other corners become growth sites. The network of branches with an intricate rectangular pattern has been shown to be multi-fractal \cite{cgd}. The growing branches never touch each other or merge together. The fact that the narrow tip continues to grow even without attachment of a visible fluid blob, indicates that in the apparently bare regions very small fluid blobs are present, which are not visible at this magnification.

\begin{figure}[ht]
\begin{center}
\includegraphics[bb=0cm 0cm 30cm 30cm, scale=0.4, angle=0]{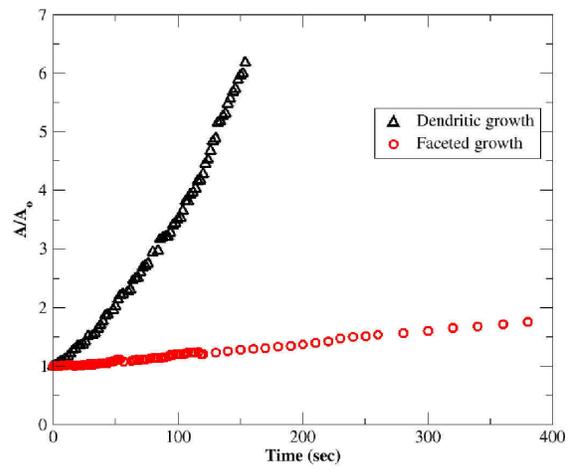}
\end{center}
\caption{Increase in area relative to a reference area as function of time are shown for a typical faceted crystal (red circles) and for a typical dendrite branch (black triangles). Compact crystal growth is always found to be slower than dendritic growth. } \label{dendrite-crystal}
\end{figure}

Typical growth rates of the large rectangular crystal and the fine dendritic structure are compared in Figure(\ref{dendrite-crystal}). The crystal aggregates are photographed at certain time intervals and grey-scaled to black and white images. The area normalized by an initial reference area is measured using imageJ software and plotted as a function of time. The rate of the large crystal growth, which is nearly linear, (a) is seen to be much slower compared to dendritic growth (b). The exact growth rates may vary from one crystal to another, but in general large faceted crystals grow more slowly compared to the dendrites.

\subsection{Effect of varying RH and salt concentration}
\begin{figure}[ht]
\begin{center}
\includegraphics[bb=0cm 0cm 30cm 30cm, scale=0.6, angle=0]{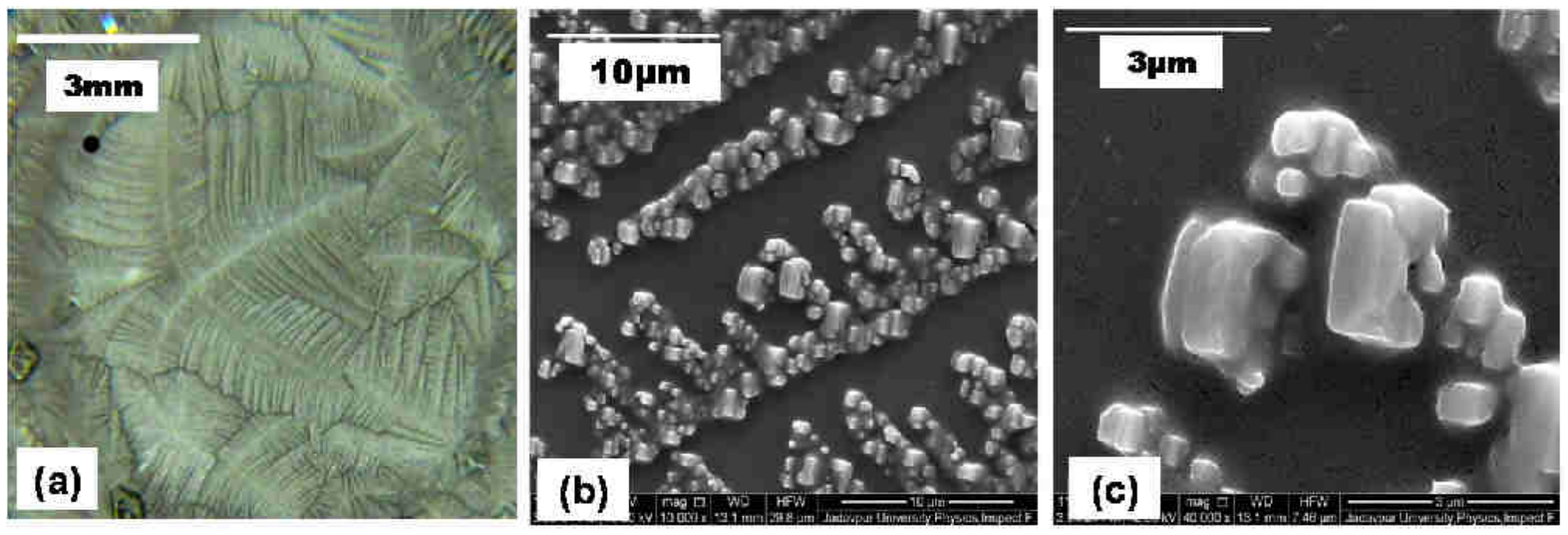}
\end{center}
\caption{Details of `ferning' (a) as seen in SEM images with lower and higher resolution (b) and (c) respectively. } \label{SEM-dendr}
\end{figure}

As RH is changed from 35\% to 45\%, the evaporation rate decreases, as seen from monitoring the weight of the drying sample with time as shown in fig(\ref{vary-rh}). The samples dried under higher RH, dry slowly and large regular crystals are seen to grow (fig \ref{vary-rh}(a),(b)). When RH is low, well-defined regular crystals are not formed in the relatively rapidly drying samples, (fig\ref{vary-rh}(c),(d)). In the background `ferning' patterns \cite{solis} are formed, rather than the dendrites with strictly rectangular symmetry. Finer details of a typical `ferning' pattern figure(\ref{SEM-dendr}a), at resolutions in the range 10 to 3 $\mu m$ are shown in figures(\ref{SEM-dendr}(b), (c)).

\begin{figure}[ht]
\begin{center}
\includegraphics[bb=0cm 0cm 30cm 30cm, scale=0.4, angle=0]{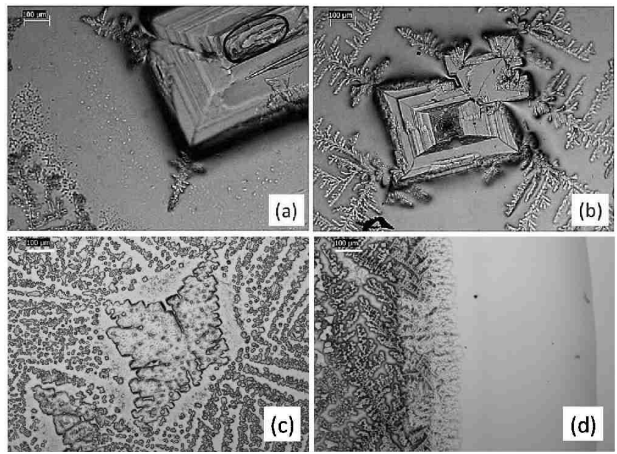}
\end{center}
\caption{Pattern formation for salt concentration 0.03 M in (a) and (b)and lower concentration 0.01 M in (c) and (d).(a) and (b) show that dendritic growth starts preferably from corners of a large crystal, (c) shows irregular growth for lower concentration and (d) shows the complex morphology near the outer periphery. The scale bars in all figures are 100$\mu$m. } \label{corner}
\end{figure}

Changing the salt concentration also causes significant changes in the pattern. For a higher salt concentration of 0.03 M NaCl, large regular crystals appear with ferns in the background (fig \ref{corner}(a),(b)). For a low salt concentration of 0.005 M mostly ferning patterns are seen with some irregular aggregates (fig \ref{corner}(c),(d)).

We develop a simulation program which takes into account the observations from the video. This program is an improved version of the simulation reported in \cite{abhracolsua}. In the earlier version only dendritic growth was considered and the evaporation process was not taken into account. In the present version, a crossover between the two growth processes is demonstrated. The kinetics is more realistic, with the differences in chemical potential between crystal and gel states taken into account while choosing preferred growth sites on the growing seed crystal. A percolation phase transition determines the crossover between the two types of growth. Details of the algorithm follow in the next section .

\subsection{Results of `incomplete drying' experiments}

From the appearance of the final dried droplet it seems as if a transparent layer of gelatin deposits on the substrate first and salt aggregates form from the remaining aqueous NaCl solution on top of it. However, the situation is not so simple, as demonstrated by the `incomplete drying' experiments.
The residue left on the glass was examined by EDAX to determine if only gelatine was deposited or some other additional constituents were present in the early deposit.  EDAX showed the presence of Cl and a slight amount of Na, in addition to C and O, the elements present in gelatine. So the layer initially deposited contains some NaCl in addition to gelatine. The still fluid layer forms a film on top of this initially deposited layer.

\section{The simulation algorithm} \label{simulation}
The two-dimensional simulation algorithm for the growing crystals, exhibits a realistic cross-over from uniform growth of large crystals to dendritic growth at a later stage of evaporation. Uniform growth along the interface of growing crystals is the dominant mechanism of growth when there is sufficient solution in the system such that the growing crystal seed is always surrounded by the solution.  It is assumed that the speed of diffusion is much greater than the speed of evaporation. This ensures that the change in salt concentration that happens due to evaporation of the solvent or the process of crystallization, is redistributed immediately in the body of the connected liquid/gel system. With evaporation, the continuous liquid/gel body breaks up into islands as more and more voids are created.
Ultimately, the voids start joining and their number starts decreasing till a single system spanning void forms outside the growing crystal.
Diffusion of solute through the liquid/gel body practically ceases at this stage. This signals the cross-over from uniform growth mechanism to the dendritic growth mechanism. In this regime, electrostatic forces are assumed to be responsible for crystal growth. This force leads to attraction of the isolated fluid blobs to the growing dendrite. The transition from the void-in-gel to gel-in-void picture can be elegantly represented by introducing a topological concept - the Euler number. This has been used in various situations involving a percolation-like transition, e.g. in crack networks \cite{vogel,sukhi-arxiv}.
 The simulation algorithm is similar to the steps outlined in \cite{book-kassner} and proceeds as follows.

\begin{enumerate}

\item{A square lattice  is placed within a circular boundary of radius 70 units. Three types of sites are possible - crystal $C$, gel $G$ and void $V$. In the beginning, the $G$ sites contain a certain concentration $c_0$ of salt, besides gelatin and water. $C$ sites are assigned a concentration  1.
 A square seed of size $4 \times 5$ is placed at the center to initiate crystal growth, these are the only $C$ sites to start with. The remaining sites are all assumed to be initially occupied by the gel containing a certain concentration of salt, these are the $G$ sites. Crystal growth is allowed only at nearest or second nearest neighbour sites of a $C$ site.
 }
\item{Evaporation may occur at any $G$ site with a probability $p_{e}$.
Upon evaporation, a $G$ site is converted into a $V$ site. The solute left behind at the site is distributed uniformly to all the connected $G$ sites. In the case of an isolated $G$ site, the solute gets deposited after evaporation of the solvent. However, the concentration of salt being insufficient for crystallization, it is deposited in amorphous form.} 
 \item{The possibility of uniform crystal growth is checked at this step. The droplet system is scanned for all $G$ sites neighbouring $C$ sites. These can convert to $C$ sites with a growth probability  decided by \cite{book-kassner}
\begin{equation}
p_{uniform}=\frac{1}{{1+exp(-\Delta E/kT)}}
\label{p_uni}
\end{equation}
Here $\Delta E$ is the change in energy that occurs due to crystal growth at a site and is given by
\begin{equation}
\Delta E = 2[(1-n_x)E_x + (1-n_y)E_y] +\Delta\mu
\end{equation} 
$k$ and $T$ are the Boltzman constant and the absolute temperature of the system, $n_x$ and $n_y$ are the number of crystal interfaces that a probable growth site encounters in the $x$ and $y$ directions respectively, $E_x$ and $E_y$ are the surface energy per unit area along the same two directions, and $\Delta\mu$ is the change in the chemical potential due to the phase transition from liquid state to     crystal state. In this simulation we have assumed that the crystal has isotropic surface energy values , i.e. $E_x = E_y$. 
 
  When the $G$ site converts to $C$, it is assumed that the salt in $G$ contributes to the growing crystal, while the solvent content is  distributed uniformly over neighbouring $G$ sites. }

\item{The connectivity of the $G$ sites is checked and all clusters of $G$ sites are counted and labelled. To start with, there is one system spanning $G$ cluster. As evaporation continues, isolated clusters of $G$ sites surrounded by $V$ sites appear. These are identified and labelled. The total number of $G$ site clusters $N_{G}$(t) at any time step $t$ is counted.}

\item{The clusters of $V$ sites that are completely surrounded by $G$ sites are identified and labelled and their number $N_{V}$(t) at any time step also noted. In the beginning, this number is zero, but as evaporation proceeds, this number $N_{V}$(t) increases. After a while,  $N_{V}$(t) starts decreasing as the void clusters start joining with each other and their number is thus reduced.
When $G$ sites no longer form a system-spanning cluster, the $V$ sites must do so, since the system is 2-dimensional. With increasing time steps, the number $N_{G}$(t) increases while the number $N_{V}$(t) decreases. }

\item{At every time step we calculate the Euler number $\chi(t)$ which is defined by

\begin{equation}
\chi(t) = N_{V}(t)- N_{G}(t)
\label{chi}
\end{equation}

The minimum of $\chi(t)$ identifies the `percolation threshold' of voids. This signals the onset of the second mechanism for crystal growth, i.e., through attractive electrostatic forces.  
Before the percolation threshold can be attained, uniform crystal growth and evaporation continue alternately.}

\item{Once the `percolation threshold' for voids is reached, growth through electrostatic forces is found to dominate over uniform crystal growth. 
 The isolated $G$ clusters are attracted towards the $C$ cluster with a force that is assumed to vary inversely as the cube of the distance between them. So a dipole-dipole type
attraction is assumed, without going into the microscopic origin of the attractive force.  A particular $G$ site attaches itself to one of the faces of the $C$ site with a probability that is given by
\begin{equation}
p_{elec}=(\Delta c/r^3)(1/S^{b})
\label{p_el}
\end{equation}

  where $\Delta c$ is the concentration difference between the sites $G$ and $C$, $r$ the distance between them, $S$ is the number of $G$ sites counted upto the nearest and second nearest neighbour sites with respect to the face and
 $b$ is a parameter. The $1/S$ dependence of $p_{elec}$ in the above equation represents the curvature of the growing crystal at the growth site. The parameter $b$ lends the weight of the curvature to the dynamics. A higher positive $b$ value will signify that more exposed growing tips of crystal are preferred as growth sites over others.
  }  
 
\item{At every time step, the total area $A$(t) of the growing crystal is measured by counting the number of $C$ sites and this is normalised by the initial area $A_0$ of the crystal seed at time $t=0$.}
\end{enumerate}

As long as the system is below the `percolation threshold' of the void clusters, a single time step skips the $7th$ step of the above procedures. Above the `percolation threshold', 
all the steps enumerated above constitute a single `time step'. The net mass of the solute is conserved at every time step.\\

\subsection{Comparison of experiment and simulation}
\begin{figure}[ht]
\begin{center}
\includegraphics[bb=0cm 0cm 30cm 30cm, scale=0.4, angle=0]{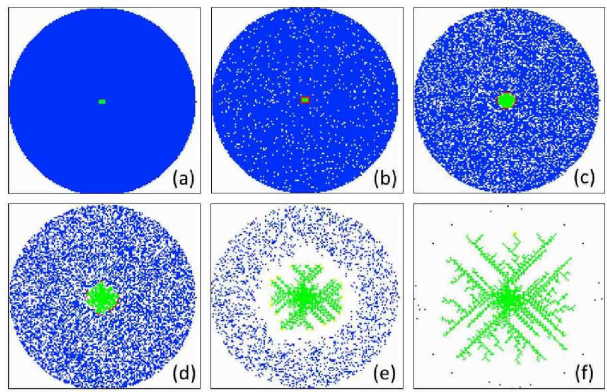}
\end{center}
\caption{ Simulating the growth from a rectangular seed. (a) shows the initial rectangular seed at the center of a fluid layer, shown in blue. (b)-(c) show uniform, faceted growth, new growth sites are shown in red. White spots in the blue fluid are voids, i.e. sites where the fluid has evaporated. (d) is the transition time with dendritic growth starting. (e) and (f) show dendritic growth, the voids now span the system, with the fluid blobs isolated.  } \label{sim-growth}
\end{figure}

Figure (\ref{sim-growth}) shows successive stages of the simulated growth with the parameters - $T$ = 1, $k$ = 1, $E_x$ = 1.10$^{-5}$, $D_m$ = 1.10$^5$, $c_0$ = 0.3, $p_{evap}$ = 0.05 and $b$ = 7. Starting with a rectangular seed in (a), initially there is uniform growth at the periphery. Evaporation creates the white dots, which are the voids in (b) and (c). The growing boundary sites following the first mechanism are shown in red, while the growth sites where the second mechanism is followed are shown in yellow. In the following time-step all red and yellow sites are coloured green identifying the growing crystal. 
In later time steps when there is no more fluid adjoining the crystal, dendritic growth starts in (d). In (e-f) dendritic growth predominates and most of the fluid has evaporated.

\begin{figure}[ht]
\begin{center}
\includegraphics[bb=0cm 0cm 30cm 30cm, scale=0.6, angle=0]{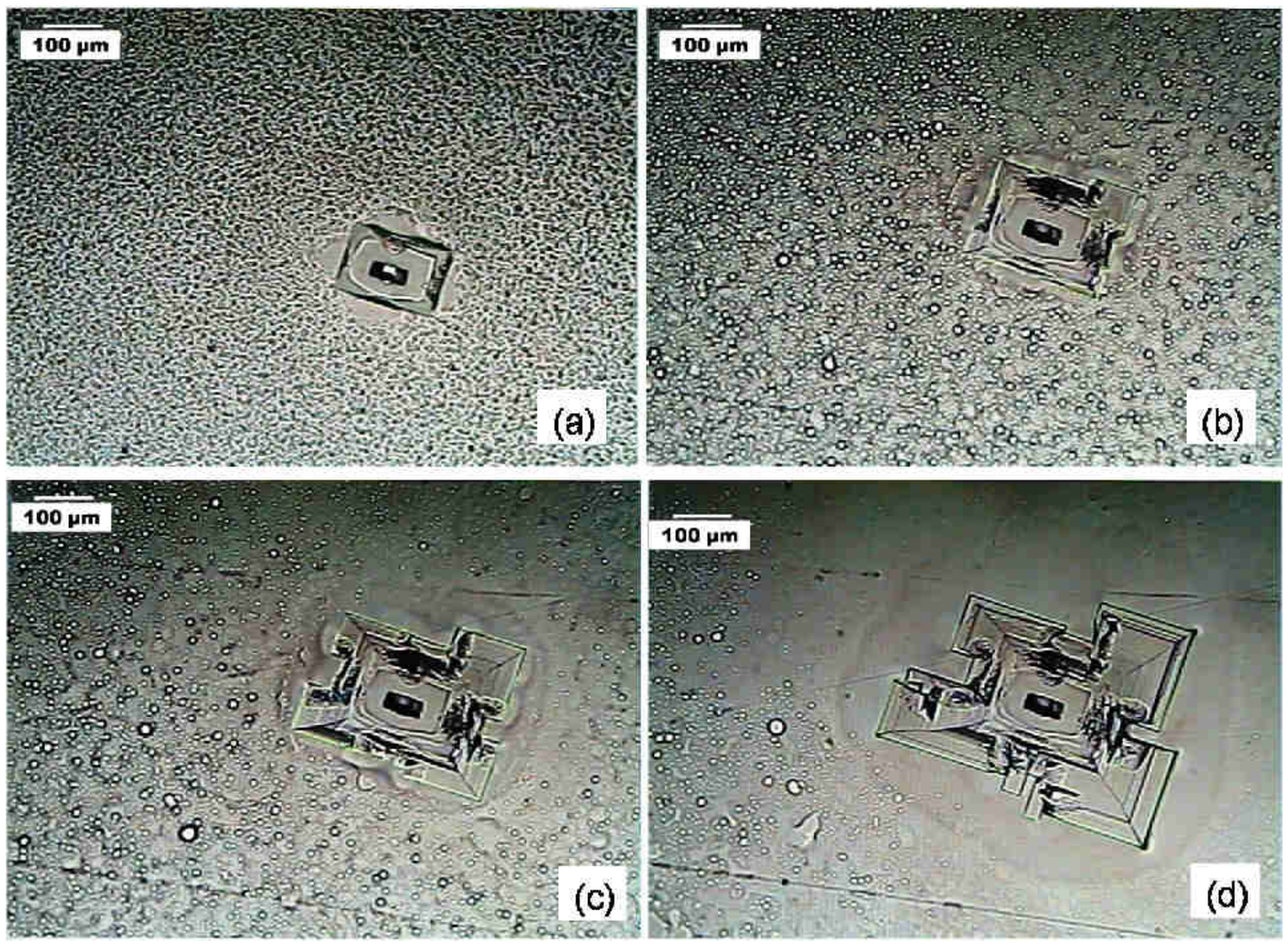}
\end{center}
\caption{The sequence (a) to (d) shows homothetic growth of the larger crystals at time (a) 0 s, (b) 5 min 14 s, (c) 7 min 15 s and (d) 10 min 15 s respectively . In (a) the gel, which contains many bubbles surrounds the rectangular seed crystal. A depletion region is seen to develop around the crystal in the subsequent figures. } \label{facet-gr}
\end{figure}

\begin{figure}[ht]
\begin{center}
\includegraphics[bb=0cm 0cm 30cm 30cm, scale=0.4, angle=0]{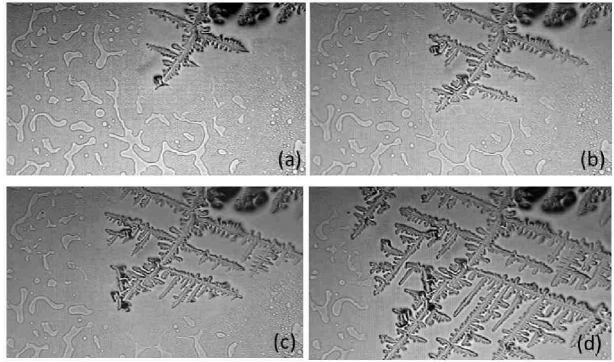}
\end{center}
\caption{Dendrite growth is shown after the `percolation' transition from connected fluid to connected voids at time (a) 0 s, (b) 1 min. 39 s, (c) 2 min. 54 s and (d) 6 min 23 s respectively. Isolated blobs of fluid appear lighter in the snapshots. The large elongated fluid blob in the lower-central part of (a), is seen to be completely swallowed up by the growing dendrite tip in (c). Larger blobs are incorporated in the dendrite as a whole and form a diamond shaped node in the dendrite branch.  The length of the rectangular panels shown in (a)-(d) is 1.73 mm.} \label{dendrite-gr}
\end{figure}

Comparing fig(\ref{sim-growth}) showing simulated growth with growth in the real system figs(\ref{facet-gr},\ref{dendrite-gr}) we see that the model  reproduces the real process quite well. Initially growth proceeds regularly along the boundary of the seed, leading to compact crystallites with straight boundaries, thus the simulated growth in fig(\ref{sim-growth}a-c) resembles the patterns formed in fig(\ref{facet-gr}). At a certain point the real system shows a change in the growth process, with dendritic growth taking over figs(\ref{corner},\ref{dendrite-gr}). We assume this to be the point when a connected layer of gel with solvent, no longer envelops the seed crystal. Our simulation leads naturally to such a situation, as seen in figs(\ref{sim-growth}d,e). The onset of the transition corresponds to the minimum in the Euler number as shown in fig(\ref{chi}). The transition from isolated void pockets in fluid to isolated fluid pockets on a substrate is also observed in the experiment as we see in fig(\ref{transition}).

\begin{figure}[ht]
\begin{center}
\includegraphics[bb=0cm 0cm 30cm 30cm, scale=0.4, angle=0]{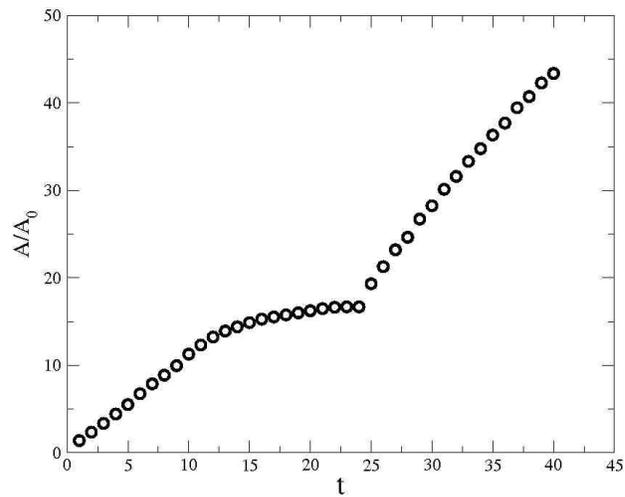}
\end{center}
\caption{Simulated aggregation from an initial rectangular seed is demonstrated by plotting the area/(initial area) against number of time-steps. Faceted uniform growth takes place initially, saturating after some time. Then faster dendritic growth starts. The two growth regimes can be clearly differentiated.} \label{growthn}
\end{figure}

The faster monotonic growth rate for dendrite growth in the simulation fig(\ref{growthn}) following  the slower growth of the large crystal is also in qualitative agreement with the measured growth rates fig(\ref{dendrite-crystal}). The simulated curve is shown as a continuous transition as it is measured over the whole system with one seed, in the real experiment faceted growth of a seed and the dendritic growth of a branch had to be measured separately. However the difference in growth rates for the large crystal and the dendrite are evident in both cases.\\

\subsection{Changing the drying conditions }

\begin{figure}[ht]
\begin{center}
\includegraphics[bb=0cm 0cm 30cm 30cm, scale=0.4, angle=0]{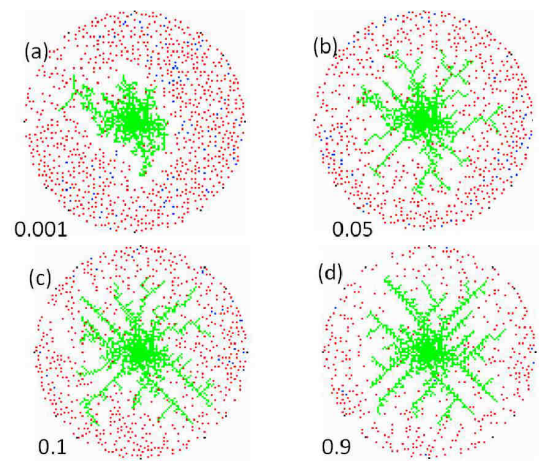}
\end{center}
\caption{The effect of varying salt concentration in simulated aggregation from an initial rectangular seed is demonstrated. The salt concentration corresponding to each pattern are given in the figure.  } \label{c-var}
\end{figure}

Changing the parameters representing salt concentration $c_0$ and the drying rate $p_{evap}$ in the simulation also produces realistic changes in the morphology. When the salt concentration is too low, large faceted crystals cannot grow due to a dearth of salt surrounding the seed fig(\ref{c-var}).

\begin{figure}[ht]
\begin{center}
\includegraphics[bb=0cm 0cm 30cm 30cm, scale=0.4, angle=0]{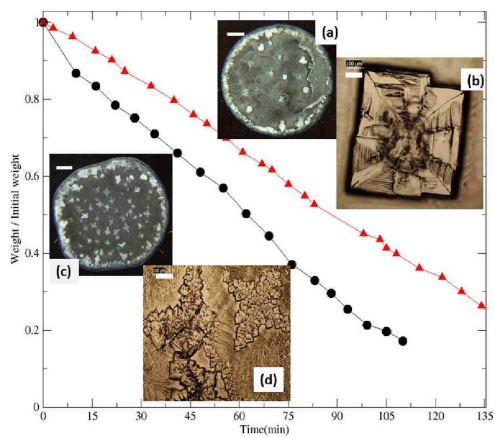}
\end{center}
\caption{Evaporation rate is lower for higher RH=45\% (red triangles) and higher for RH=35\% (black circles) as seen from the weight variation of the droplets with time. Corresponding patterns on dried droplets are shown in (a), (b) for RH = 45\% and (c), (d) for RH = 35\% at for different length scales. The scale bar shown in (a) and (c) is 1.8 mm, for (b) and (d) the scale bar is 100 $\mu m$. } \label{vary-rh}
\end{figure}

\begin{figure}[ht]
\begin{center}
\includegraphics[bb=0cm 0cm 30cm 30cm, scale=0.4, angle=0]{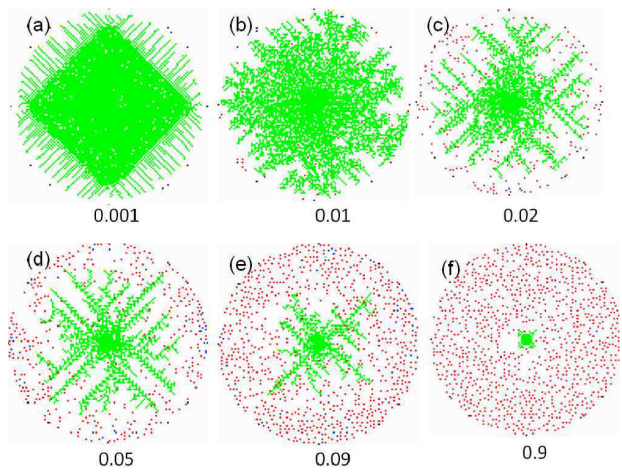}
\end{center}
\caption{The effect of varying evaporation rate in simulated aggregation from an initial rectangular seed is demonstrated. The evaporation rates corresponding to each pattern are given in the figure. }
\label{rh-var}
\end{figure}

Our simulations show that with slow evaporation, i.e. small $p_{evap}$, large faceted crystals form and dendrite growth is much reduced. With higher evaporation rate, the dendrites grow preferably from corners of the large crystal as shown in fig.(\ref{rh-var}). This is in agreement with the real growth fig(\ref{vary-rh}(b),(d)). 

The simulation shows that as salt concentration increases irregular growth fig(\ref{c-var}(a)) changes over to regular growth. The agreement between experiments and simulation is qualitatively valid, however it is difficult to control humidity, temperature and salt concentration simultaneously in our present experimental set up.

The effect of varying the evaporation rate has also been investigated both experimentally (fig \ref{vary-rh}) and through simulations. Since the ambient temperature, humidity as well as the salt concentration all determine the drying rate, we have measured the weight loss of the sample directly. Figures \ref{vary-rh}(c),(d) show that for fast drying small irregular aggregates are scattered quite densely over the drop, while for slow drying large regular crystals have enough time to grow (fig \ref{vary-rh}(a),(b)) before the solvent dries out. This is supported by the simulation results shown in fig.(\ref{rh-var}).

In addition to the two growth mechanisms - facet growth and dendritic growth, a third type of formation, `ferning' is observed under certain conditions (fig.\ref{SEM-dendr}). In this work we have not probed the details of this mechanism and plan to do this in the future.

There are several factors affecting pattern growth like temperature and relative humidity, the salt concentration also changes the surface tension, which in turn affects evaporation rate \cite{moucolsua}. We plan to try a more precise specification of drying conditions in future experiments. The effect of temperature in the simulation has not been taken into account in this report, a simultaneous systematic variation of all parameters in the simulation will also be carried out.

\section{Discussion}
Study of dried droplet patterns is no longer only an interesting academic exercise.
The typical morphological features of the dried droplet are now being put to use in novel devices and applications. The coffee-ring effect \cite{deegan}, is used by Layani et al. \cite{layani} to fabricate transparent and flexible conducting sheets. Arrays of very thin interconnected rings composed of metallic nano-particles are produced by drying droplets on a plastic sheet. They make the sheet conducting and since most of the sheet has no metallic deposit, it is also transparent. Carbon nano-tubes have been used similarly by Shimoni et al. \cite{shimoni}.

Fern-like crystal growth in dried body fluids like tear drops helps in diagnosing diseases like `dry eyes', while `ferning' in dried cervical fluid is used to monitor human fertility \cite{solis}. Typical crack patterns in dried drops of blood or plasma helps in diagnosing diseases related to blood \cite{brutin}, while patterns from dried drops containing starch  like wheat can be used for quality control of food grains \cite{wheat-grain}. Bacteria have also shown a coffee-ring effect in recent work \cite{bacteria}. When the medium is a gel, convection, leading to the coffee ring is usually suppressed to some extent, so an added salt may crystallize in a greater variety of morphological patterns. 

\section{Conclusions}

We have presented experiments on gelatin and salt in water, with a well-defined cross-over from the compact crystal morphology to a dendritic pattern. Our simulation algorithm, though very simple, can reproduce this cross-over from one regime to another as evaporation proceeds. Evaporation is explicitly taken into account in this model and it produces a time varying pattern of interspersed fluid and void sites.  An interesting finding is that an Euler number can be defined for the pattern, which has a minimum very close to the transition point between two regimes of aggregation.
In the first regime of compact faceted growth, the growth probability follows standard prescriptions \cite{book-kassner}. In the subsequent dendritic regime, experiments show finite fluid blobs  moving towards the growing dendrite tip. Since non-uniform charge distributions are present in the complex fluid containing ions and colloids, the  attraction may be assumed to be of electrical origin \cite{pimpinelli}. However, the exact form of the interaction needs further investigation. For the present we have assumed it have an inverse cubic variation with distance.

 In this preliminary work, we deal with crystal growth from one seed, but we have plans to extend this to cases of multiple, simultaneously growing seeds as seen in the real experiment. Studying the effect of varying the parameters, e.g. controlling the evaporation rate, salt concentration etc. and comparing with experiments are planned for future work.

\section{Acknowledgment}
This work is funded by DST, Govt. of India through research project No. SR/S2/CMP-127/2012. MDC is grateful to CSIR, India for award of a Senior Research Fellowship. ST thanks the Northwestern Polytechnical University for hospitality and arranging a visit to Xian.  Duyang Zang, Xiaopeng Chen, Guangyin Jing and Yuri Tarasevich are gratefully acknowledged for stimulating discussion and helpful suggestions. Authors thank DST Government of India for funding SEM facility at Jadavpur University (configuration no. QUO-35357-0614) through FIST-2.

\newpage

\newpage

\newpage

\newpage
\newpage
\newpage

\newpage

\newpage
\begin{figure}[ht]
\begin{center}
\includegraphics[bb=0cm 0cm 30cm 30cm, scale=0.4, angle=0]{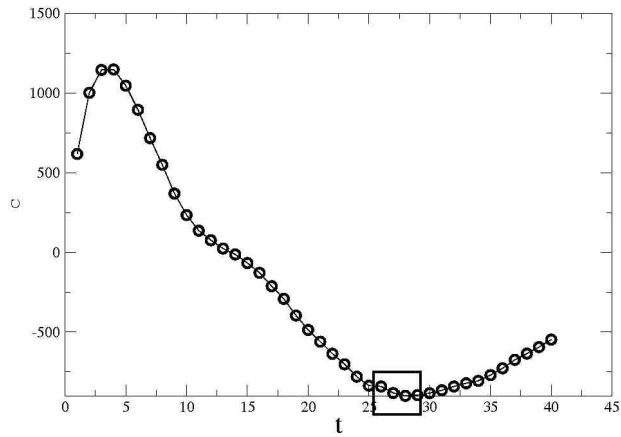}
\end{center}
\caption{Variation in the Euler number $\chi(t) = N_{V}(t)- N_{G}(t)$ with time (i.e. number of time steps) as evaporation proceeds. A box marks the region where $\chi$ reaches a minimum, before starting to grow again. The minimum in $\chi$ corresponds closely to the time when dendritic growth takes over from uniform growth, as seen in figure(\ref{growthn}). } \label{chi}
\end{figure}

\newpage

\newpage

\newpage

\end{document}